\documentclass[11pt]{article}

\date{}

\usepackage{gensymb}
\usepackage{textcomp}
\usepackage{adjustbox}
\usepackage[utf8]{inputenc} % allow utf-8 input
\usepackage[T1]{fontenc}    % use 8-bit T1 fonts
\usepackage{url}            % simple URL typesetting
\usepackage{booktabs}       % professional-quality tables
\usepackage{amsfonts}       % blackboard math symbols
\usepackage{nicefrac}       % compact symbols for 1/2, etc.
\usepackage{microtype,nicefrac}      % microtypography
\usepackage{xspace}
\usepackage{natbib}
\usepackage{comment}
\usepackage{amsmath}
\usepackage{graphicx}
\usepackage{rotating}
\usepackage{amssymb}
\usepackage{autobreak}
\usepackage{mathtools}
\usepackage[dvipsnames]{xcolor}
\usepackage{bbm}
\usepackage[ruled,vlined, linesnumbered]{algorithm2e}
\usepackage{algorithmic}
\usepackage[parfill]{parskip}
\usepackage{color}
\usepackage[urlcolor=blue,citecolor=black,linkcolor=black]{hyperref}
\usepackage{amsmath} % provides the "align" environment
\usepackage{graphicx}
\usepackage{subfig}
\usepackage{amssymb} 
\allowdisplaybreaks

\title{Imaging the Northern Los Angeles Basins with Autocorrelations}
\author{Caifeng Zou and Robert W. Clayton \\\\
   \textit{Seismological Laboratory, California Institute of Technology, Pasadena, CA \emph{91125}, USA }\\
  }

\begin{document}

\label{firstpage}

\maketitle

\begin{abstract}
We show reflectivity cross-sections for the San Gabriel, Chino, and San Bernardino basins north of Los Angeles, California determined from autocorrelations of ambient noise and teleseismic earthquake waves. These basins are thought to channel the seismic energy from earthquakes on the San Andreas Fault to Los Angeles and a more accurate model of their depth is important for hazard mitigation. We use the causal side of the autocorrelation function to determine the zero-offset reflection response. To minimize the smoothing effect of the source time function, we remove the common mode from the autocorrelation in order to reveal the zero-offset reflection response. We apply this to 10 temporary nodal lines consisting of a total of 758 geophones with an intraline spacing of 250-300 m. We also show that the autocorrelation function from teleseismic events can provide illumination of subsurface that is consistent with ambient noise. Both autocorrelation results compare favorably to receiver functions.
\end{abstract}

\section{Introduction}
Sedimentary basins with low velocities and densities can trap and amplify seismic waves from earthquakes, resulting in stronger and more prolonged ground shaking. This can pose a threat to the urban infrastructure on top of these basins. The San Gabriel, Chino, and San Bernardino basins northeast of Los Angeles (LA) are located in a seismically activate area near the southern segment of the San Andreas Fault (SAF), which has been identified as a major source for large earthquakes. These sedimentary basins can channel the earthquake energy from the SAF toward the metropolitan Los Angeles. 

Numerical simulations including TeraShake \citep{olsen2006strong,olsen2008terashake2}, ShakeOut \citep{graves2008broadband}, and CyberShake \citep{graves2011cybershake} have shown intense ground motions in the densely populated Los Angeles region from large ruptures on the southern SAF. However, the accuracy of these simulations relies heavily on the assumed velocity models. \cite{denolle2014strong} used ambient noise cross-correlations to model virtual earthquakes on the SAF, predicting about four times stronger ground shaking in downtown LA than simulations with the current basin models. The less-constrained velocity models for the northern LA basins, which can act as a waveguide in channeling seismic energy into the city \citep{olsen2006strong}, are candidates for underestimation of the potential seismic hazards in above-mentioned simulations.

\begin{figure}
\includegraphics[width=1.0\textwidth]{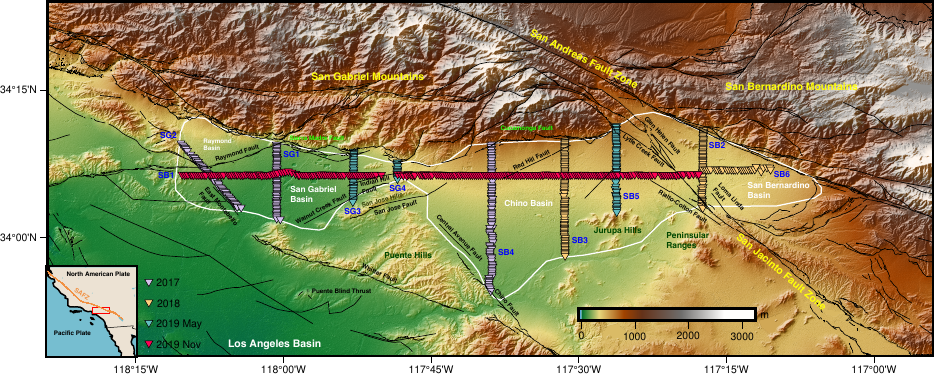}
\caption{Elevation map of the studied area denoted by the white polygon. The 10 nodal lines in the BASIN survey are delineated with triangles in different colors grouped by the four deployment times between 2017 and 2019. Faults in the area are plotted with black lines, according to the \href{https://www.usgs.gov/natural-hazards/earthquake-hazards/faults}{U.S. Geological Survey database}.}
\label{studied_area}
\end{figure}

The Basin Amplification Seismic Investigation (BASIN) project was designed to better understand the structures and material properties of the northern LA basins, to improve earthquake hazard simulations and assessments in the Los Angeles metropolitan area. The BASIN project deployed short-term dense nodal arrays in the San Gabriel, Chino, and San Bernardino basins (Figure \ref{studied_area}), where active source surveys are limited and broadband stations are sparse. Previous studies using the data from this survey include traditional frequency domain deconvolution receiver functions (RF) \citep{wang2021urban,ghose2023basin}, Bayesian array-based coherent RF \citep{wang2021urban}, a 3D shear wave velocity model using ambient noise tomography \citep{li2023shear}, and a 3D basin depth model integrating RF, gravity, and borehole data \citep{villa2023three}. Autocorrelation is another method that can be used to image structure in the crust. \cite{clayton2020imaging} showed ambient noise autocorrelations for one of the 10 nodal transects in the BASIN survey to image basements and faults as an example. Here we show the basement structure derived from autocorrelations for all the lines in the survey.

The underlying principle was proposed by \cite{claerbout1968synthesis}, who showed that the zero-offset reflection response of a horizontally layered acoustic medium can be retrieved from the autocorrelation of the transmission response. The theory was later generalized by \cite{wapenaar2003synthesis} to arbitrary heterogeneous, elastic media. Teleseismic waves arrive at the sensors on top of sedimentary basins with steep incidence and reverberate between discontinuities. The resulting repeating two-way travel times between the free surface and subsurface interfaces can be extracted through autocorrelation. Autocorrelating teleseismic waves has been used to image crustal seismic discontinuities \citep{ruigrok2012global,sun2016receiver,pham2017feasibility,viens2022imaging}. 
A number of studies \citep{tibuleac2012crust,zhan2014ambient,taylor2016crustal,oren2017seismic,romero2018mapping,clayton2020detailed} have also applied the autocorrelation technique to ambient noise, which is generated by diffuse, directionless sources such as the coupling of ocean waves.

In this study, we create auto-correlation functions (ACF) from both teleseimsic waves and ambient noise to image the San Gabriel, Chino, and San Bernardino basins. Because there were only a few earthquakes recorded during the node deployment period, we select one single event for each nodal transect to create earthquake ACFs rather than stack multiple events. Even so, the earthquake ACFs produce results consistent with the noise ACFs. The autocorrelations capture geologically reasonable basin structures that are compared with previous receiver function results. We develop better understandings of the studied basins, which play a significant role in accurately simulating the ground shaking in the greater LA region caused by earthquakes from the SAF.

\section{Methods}
\subsection{Retrieving reflectivity through autocorrelation}
Equation \ref{claerbout} describes the relationship in the $z$-transform domain, where $A(z)$ is the autocorrelation function, $T(z)$ is the transmission response, $R(z)$ is the zero-offset reflection response, and $S(z)$ is the source function. 
\begin{equation}
\begin{aligned} 
 A(z)=T(z)T(\frac{1}{z})=(1+R(z)+R(\frac{1}{z}))S(z).
\label{claerbout}
\end{aligned}
\end{equation}
Typically $ \left| R\right|\ll 1$, so the source term dominates the ACF and needs to be removed to reveal the reflection response. Standard deconvolution is usually not effective because the source has a narrow frequency band. Instead, we use a method shown in \cite{clayton2020imaging} to estimate the source term with the average autocorrelation over the seismic profile, referred to as the common mode:
\begin{equation}
\begin{aligned} 
S_{est}(z)=\frac{1}{N} \sum_{i=1}^{N}A_{i}(z),
\label{sest}
\end{aligned}
\end{equation}
where $N$ is the number of stations along the seismic line and $A_i(z)$ is the ACF of a single station. Taking the causal side of ACFs, the reflection response is estimated by:
\begin{equation}
\begin{aligned} 
R_{est}(z)=R(z)S(z)=A(z)-S_{est}(z).
\label{rest}
\end{aligned}
\end{equation}

\begin{figure}
\includegraphics[width=1.0\textwidth]{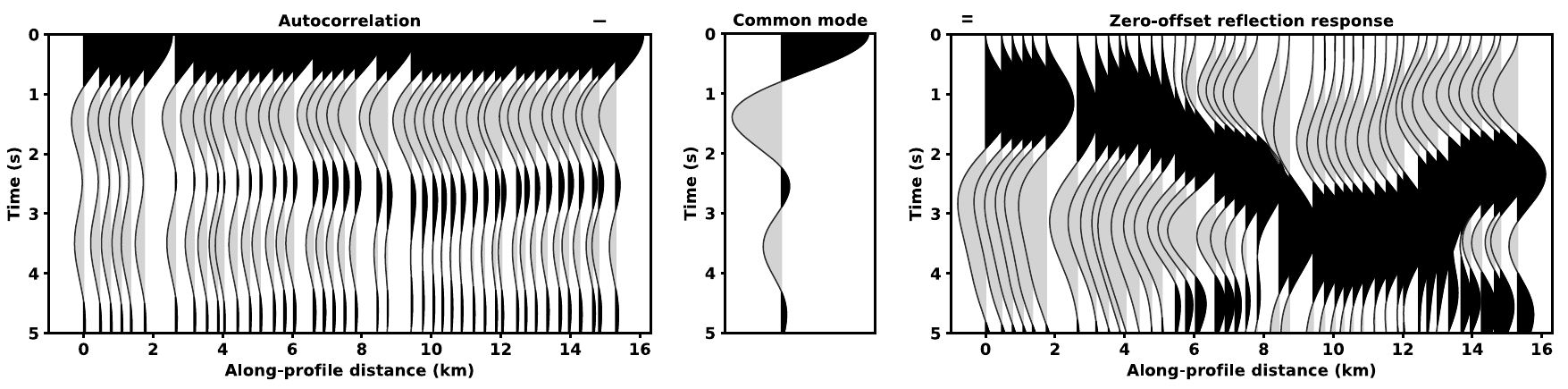}
\caption{Removing the common mode from the autocorrelation to unveil the zero-offset reflection response. The common mode is given by the average autocorrelation over the seismic profile. This example is the noise ACF taken from SG2.}
\label{rmcm}
\end{figure}

Figure \ref{rmcm} shows the effect of common mode removal, taking the noise ACF from SG2 as an example. Before removing the common mode, the source term dominates the ACF and obscures the reflectivity. By subtracting the common mode from the ACF of each station, the underlying zero-offset reflection response can be retrieved. The drawback to this approach is that horizontal geological features are also removed by this process. Fortunately most of the structures of interest (i.e. basins) are not flat.

\subsection{Processing}
\begin{figure}
\includegraphics[width=1.0\textwidth]{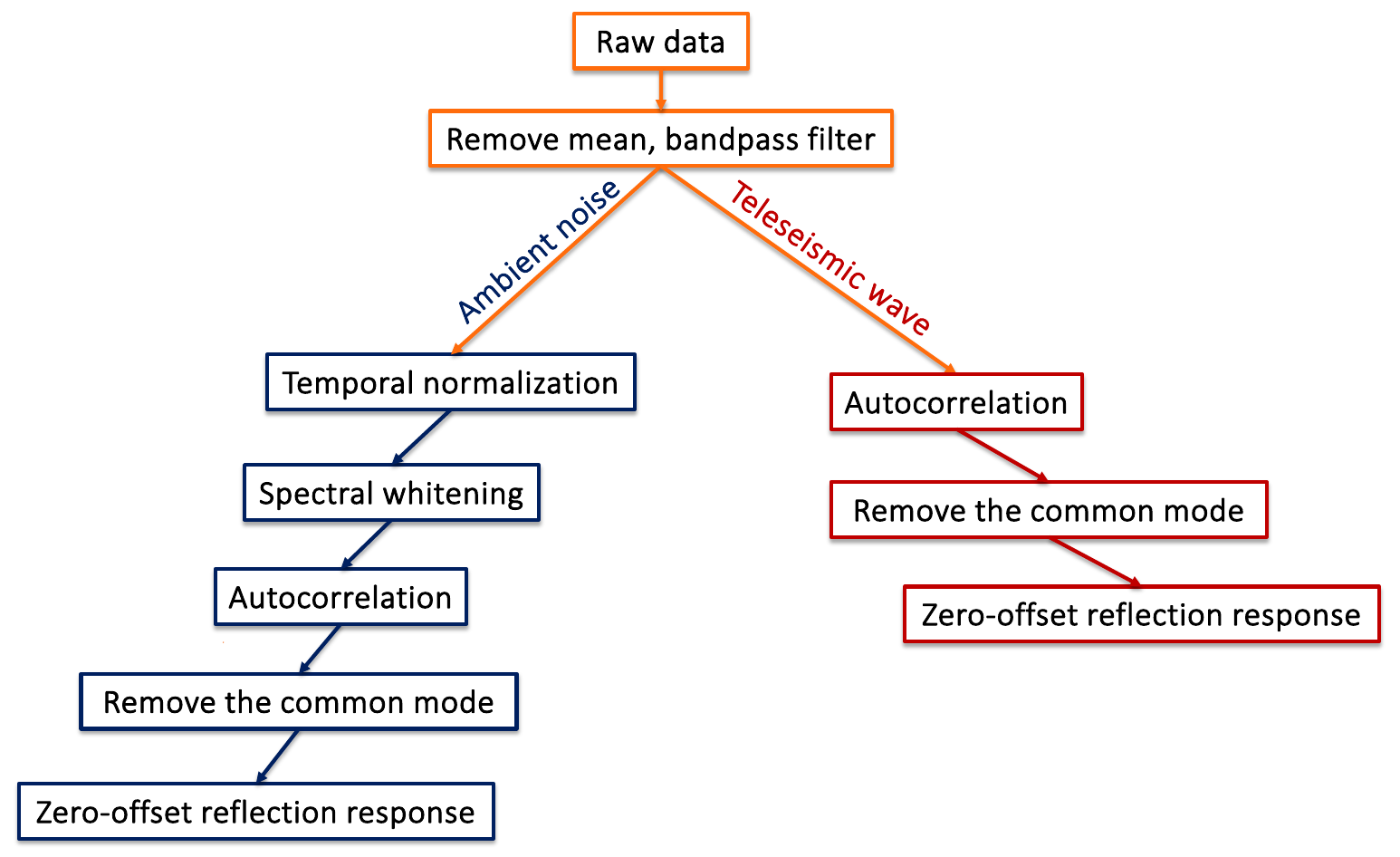}
\caption{Workflow of imaging with ambient noise and teleseismic wave autocorrelations.}
\label{workflow}
\end{figure}

We process ambient noise and earthquake data following the method proposed by \cite{bensen2007processing}, as illustrated in Figure \ref{workflow}. The two types of data share most processing steps, except for two steps that are exclusively implemented for ambient noise: temporal normalization and spectral whitening. We remove the mean of raw seismic signals in a trace-wise manner and filter the data in the $[0.1,1]$ Hz range. This frequency range has a good signal-to-noise ratio (SNR) as determined by trial and error. To prevent earthquakes from dominating the ambient noise correlations, we apply a running-absolute-mean normalization in the time domain (temporal normalization):
\begin{equation}
\begin{aligned} 
\tilde{d}_{i}=\frac{ d_{i}}{\hat{\omega }_{i}},
\label{di}
\end{aligned}
\end{equation}

\begin{equation}
\begin{aligned} 
 \hat{\omega }_{i}= \frac{1}{2N+1}\sum_{j=i-N}^{i+N}\left| \hat{d }_{j} \right|,
\label{wt}
\end{aligned}
\end{equation}
where  $d_i$ and  $\tilde{d}_{i}$ denote raw and normalized seismograms respectively, and $\hat{\omega }_{i}$ is the point-wise normalization weight, which is calculated within a window with a width of $(2N+1)$. The width used in this study is set to 5 s, which is half the maximum period of the applied filter, as recommended in \cite{bensen2007processing}. Note that the temporal weights $\hat{\omega }_{i}$ are not computed using the raw waveform data $d_j$ but on data $\hat{d }_{j}$ filtered in the earthquake reject band of $[15,50]$ s. This has proven to be necessary and effective for suppressing earthquake signals \citep{bensen2007processing,zhan2010retrieval}. Spectral whitening is conducted for ambient noise to balance the frequency band of signals and to attenuate the effect of some persistent monochromatic sources. It is the spectral counterpart of temporal normalization:
\begin{equation}
\begin{aligned} 
 \tilde{s}_{i}=\frac{ s_{i}}{\frac{1}{2N+1}\sum_{j=i-N}^{i+N}\left| s_{j} \right|},
\label{whiten}
\end{aligned}
\end{equation}
where $s_i$ is the original spectrum and $\tilde{s}_{i}$ is the whitened spectrum. We empirically set the spectral window width $(2N+1)$ to 0.5 Hz.

\begin{figure}
\includegraphics[width=1.0\textwidth]{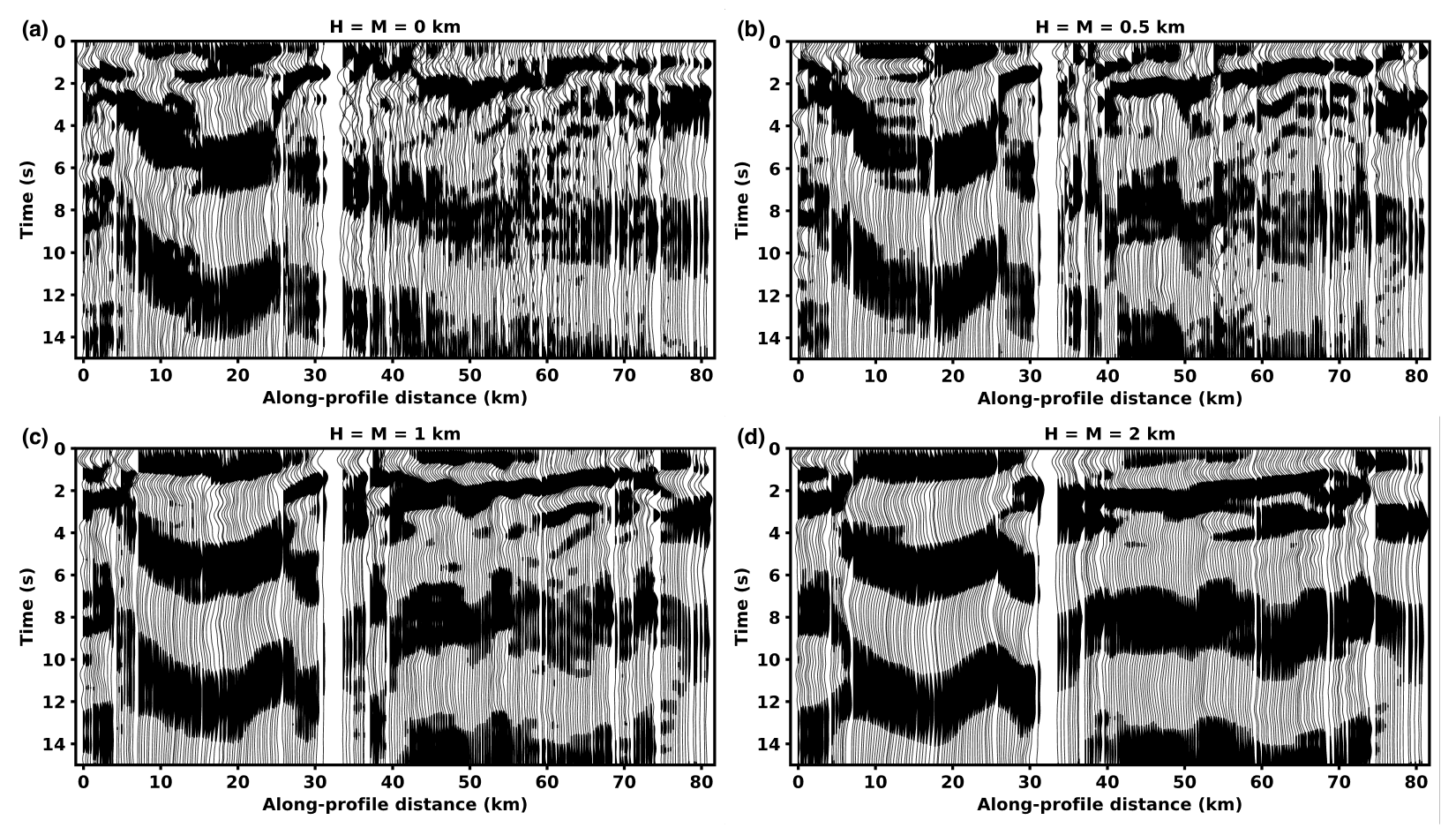}
    \caption{Effect of stacking near-zero offset cross-correlations with (a) $H=M=0$ km (pure autocorrelations), (b) $H=M=0.5$ km, (c) $H=M=1$ km, and (d) $H=M=2$ km on the imaging results with ambient noise. This example is taken from SB1.}
    \label{ccstacking}
\end{figure}

For ambient noise waveforms, we compute the autocorrelations over the node deployment period of a month approximately. We believe this correlation length is adequate because we empirically find that a correlation length of a week suffices to produce results similar to those obtained with a month. The correlation length for teleseismic waveforms is set to 50 s, covering the P/S arrivals and subsequent coda waves. Strictly speaking, autocorrelations are computed in a trace-wise manner without interactions between different stations. However, in order to enhance the SNR and robustness against individual bad points, we approximate the noise ACFs by stacking near-zero offset cross-correlations \citep{clayton2020imaging}: 
\begin{equation}
\begin{aligned} 
 A(\textbf{x})=\sum C(\textbf{x}_i,\textbf{x}_j),\left| \textbf{x}_i-\textbf{x}_j\right|\leq H,\left| \textbf{x}-\frac{\textbf{x}_i+\textbf{x}_j}{2}\right|\leq M,
\label{stack}
\end{aligned}
\end{equation}
in which $A(\textbf{x})$ is the autocorrelation for trace $\textbf{x}$ and $C(\textbf{x}_i,\textbf{x}_j)$ is the cross-correlation between traces $\textbf{x}_i$ and $\textbf{x}_j$. The sum is over all pairs of stations $(\textbf{x}_i,\textbf{x}_j)$ such that the distance between $\textbf{x}_i$ and $\textbf{x}_j$ is less than $H$ and the distance between station $\textbf{x}$ and the midpoint of the pair is less than $M$. There is a trade-off between the SNR and lateral resolution in deciding $H$ and $M$. Large values sacrifice the lateral variation for robustness. Figure \ref{ccstacking} shows an example from SB1 of how stacking with different values of $H$ and $M$ affects the final imaging results with ambient noise. In this study, we choose $H=M=1$ km for noise ACFs, which strikes a good balance between the basin shape variation and the lateral continuity. Note that the averaging over offset ($H$) is done without any normal moveout correction, in order to suppress surface waves. Even with pure autocorrelations ($H=M=0$ km), the overall features of subsurface are captured. 

For earthquake ACFs, we set $M=1$ km and $H=0$ km, which means that no offset averaging is done. Since only a small number of $M_w\geq6$ earthquakes occurred during the one-month deployment, we select one single event for each nodal transect with good signal instead of stacking the results over multiple events. The latter is more commonly adopted when sufficient events are available. The performance of teleseismic events in imaging the subsurface depends on a number of factors, including magnitude, epicentral distance, incidence angle, and data SNR. We scan all events for all lines and select the ones that produce results that are geologically reasonable. Detailed information regarding selected earthquakes is provided in Table \ref{events}. More information on the teleseismic event can be found in \cite{ghose2023basin}.

As introduced earlier, the common mode is subtracted from each trace of noise and earthquake ACFs to estimate the zero-offset reflection response. The methodology works for multiple (Z, R, T) components. We assume that ZZ, RR, and TT ACFs are dominated by P, SV, and SH reflectivity, respectively. Although the T and R components are not strictly defined at a single point, we denote the direction for teleseismic waves by the back azimuth from the station to the epicenter and for ambient noise by the direction between cross-correlation pairs.

\section{Data and Studied Area}
Figure \ref{studied_area} shows the elevation around the studied area and locates the BASIN survey. The BASIN survey deployed dense nodal lines in the northern LA basins over four different periods between 2017 and 2019. The 10 lines consisted of 758 Fairfield ZLand three-component nodal sensors spaced with 250-300 m and collected waveform data for 30-35 days. The corner frequency of the sensors is 5 Hz. The nodes were distributed over a wide urban area that does not have 3D oil company surveys. There are four lines in the San Gabriel Basin (SG1, SG2, SG3, and SG4), three lines in the Chino Basin (SB3, SB4, and SB5), and two lines in the San Bernardino Basin (SB2 and SB6). The long, west-east-trending line SB1 crosses the San Gabriel and Chino basins and ends near the western edge of the San Bernardino Basin. The number of $M_w\geq6$ earthquakes recorded by each nodal line and the information about the selected event for computing earthquake ACFs are given in Table \ref{events}. More details regarding the BASIN survey can be found in \cite{clayton2019exposing}. The sedimentary formations in the studied area result from the opening of the Los Angeles Basin during the Miocene \citep{wright1991structural}. We refer to \cite{villa2023three} or \cite{ghose2023basin} for a more detailed description of the geologic setting in the San Gabriel, Chino, and San Bernardino basins. 

\section{Results}
\begin{figure}
  \centering
  \rotatebox{270}{
    \begin{minipage}{0.98\textheight} 
      \includegraphics[width=1.0\textwidth]{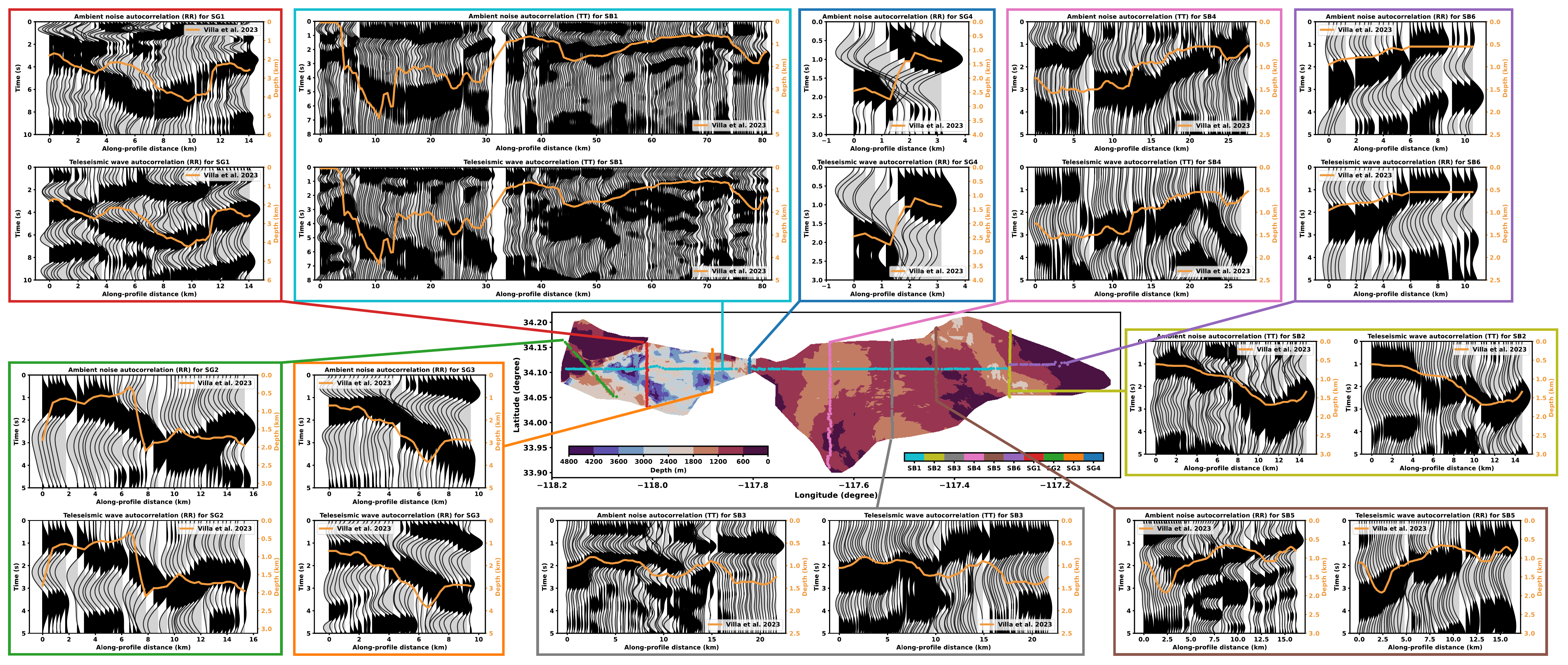}
      \caption{Imaging results of ambient noise and teleseismic wave autocorrelations for the northern LA basins, in comparison to an earlier study on the basin depth \citep{villa2023three}.}
      \label{compile}
    \end{minipage}
  }
\end{figure}

We show results for a number of lines in the survey along with some geologic interpretations (Figure \ref{detailedline}) and comparisons to receiver functions for all of the 10 seismic lines (Figure \ref{compile}). For each line we have tried all of the three components, but the horizontal (R and T) components produce consistently better (more continuous, less noisy) images than the vertical (Z) component. We attribute this to data quality due to the stronger shear component in the raw seismograms. We present the best images from ambient noise and teleseismic waveforms for each line. We compare our imaging results using autocorrelations directly to the depth of the sediment-basement interface obtained from an earlier study \citep{villa2023three}, where integrated gravity and receiver function (RF) measurements, along with borehole constraints, are used to produce a 3D depth map for the basins. We choose this study for comparison because it has incorporated the RFs computed in all of the three previous studies \citep{liu2018structure,wang2021urban,ghose2023basin} in the same area. We simplify things by focusing on the basin shapes, so no time-to-depth conversion is done on ACFs. For comparison, we plot two vertical axes: time and depth, where the velocity model of \cite{li2023shear} has been used to obtain the depth. The same amount of smoothness ($M=1$ km) is applied to both results in the comparison.

Figure \ref{detailedline}a shows the imaging results for SB1, which is the longest line in the BASIN survey crossing the San Gabriel and Chino basins from west to east and extending to the edge of the San Bernardino Basin. The ambient noise and earthquake ACFs generally align with each other, with both images showing a deeper San Gabriel Basin and a shallower Chino Basin. The lateral variation of the sediment-basement interface is consistent with the basin depth variation based on \cite{villa2023three}. This is especially true for the Chino Basin, which could be explained by that it has a more homogeneous velocity model than the San Gabriel Basin \citep{li2023shear}. We also find the strong multiple of the San Gabriel Basin delineated in red and a slab-like geologic feature delineated in blue that is not seen by other methods but has no tectonic interpretation. Figure \ref{detailedline}b shows the images for SG1 that crosses the San Gabriel Basin from north to south. It also has a deep basin, in agreement with SB1. The basin shape revealed by the time-domain ACFs generally resembles it in depth. Again, the difference in them could be explained by the strong heterogeneities of the underlying velocity model \citep{li2023shear}. In these two cases the earthquake ACFs are noisier than the noise ACFs, as only one event is used for imaging. The incidence angle of teleseimic waves through sedimentary basins is assumed to be vertical, but this is not strictly true, resulting in slightly different times.

SG2 lies near the southwest border of the San Gabriel Basin and crosses the left-lateral Raymond Fault bounding the basin to the northwest. The Raymond Fault acts as a major barrier between the Raymond Basin and the deeper San Gabriel Basin. A steep Bouguer gravity gradient is observed southeastward along SG2 over the Raymond Fault, due to the deepening basement and change of rock density, as is reflected in the basin depth variation (Figure \ref{detailedline}c). The noise and earthquake ACFs also see a remarkable deepening feature at the same location although it is not as sharply defined.

Figure \ref{detailedline}d presents consistent results from ambient noise and teleseismic autocorrelations for SG3 near the eastern boundary of the San Gabriel Basin. It shows a southward deepening of the basin after passing the (left-lateral, strike-slip) Indian Hill Fault in agreement with the depth model. The (northeast-striking, left-slip) Walnut Creek Fault separates the San Gabriel Basin from the San Jose Hills and sees a southward rise in depth (Figure \ref{studied_area}), but this is either not captured or smoothed out in ACFs.

In the north-to-south SB4 crossing the Chino Basin (Figure \ref{detailedline}e) we find a fault-like feature delineated in red. This is interpreted as the Red Hill Fault, which is a thrust fault, that is seen in an earlier study on ambient noise autocorrelations \citep{clayton2020imaging} and two studies on teleseismic receiver functions \citep{liu2018structure,ghose2023basin}. However, the feature does not appear in the comprehensive basin depth model \citep{villa2023three}, as it also incorporates gravity measurements resulting in the smoothing out of local details.

SB2 is located in the San Bernardino Basin and crosses the SAF, Loma Linda Fault, and San Jacinto Fault from north to south. As shown in Figure \ref{detailedline}f, the very shallow north end along SB2 reflects the fact that the San Bernardino Basin is bounded by the SAF to the northeast. To the south, the sediment-basement interface deepens and reaches its deepest point near the Loma Linda Fault, in agreement with \cite{ghose2023basin}. The interface turns shallower again further south till it meets the San Jacinto Fault at the end of SB2.

Figure \ref{compile} integrates the noise and earthquake ACFs for all of the 10 seismic lines in the northern LA basins for a comprehensive view. In the middle is the map of the basin depths developed by \cite{villa2023three} along with the locations of the nodal lines plotted on top. Each line points to its own images with a corresponding color. The lines SB3 and SB5 in the Chino Basin are relatively noisy but still exhibit consistency in basin shape. The shortest line in the survey, SG4 at the throat connecting the San Gabriel and Chino basins, is less than 4 km. Even so, the ACFs show an impressive agreement with the basin depth variation. The west-east-oriented SB6 line sees a shallowing of the sediment-basement interface toward the northeast end of the San Bernardino Basin bounded by the SAF.

\section{Discussion}
The autocorrelation technique can also be interpreted as a special case of the broader theory known as seismic interferometry, where the Green's function emerges from the cross-correlation of diffuse waves (e.g., ambient noise) recorded at two stations. With a single station, the cross-correlation becomes the autocorrelation, and the zero-offset reflection section is imaged. Thus autocorrelation processing is similar to cross-correlation processing, but additional care is needed for the source time function. Spectral whitening is carried out to produce a broader-band signal and reduce the imprint of the source wavelet. However, because the autocorrelation function is naturally zero-phase, perfect (or one-bit) whitening will lead to a delta function at zero lag and conceal the later reflections. Therefore, we do spectral whitening within the passband of $[0.1,1]$ Hz in a running average manner with a window of 0.5 Hz. Even so, the source term spiked at 0 s still has dominant energy in the ACF. While tapering the first few data points \citep{viens2022imaging,pham2017feasibility,mroczek2021joint} and automatic gain control (AGC) \citep{oren2017seismic} have been used to suppress the source imprint, we employ a common mode removal technique that approximates the source term with the average ACF over the transect, which is then subtracted from each trace to reveal the reflections. A possible negative side effect of this method is that horizontal interfaces will be attenuated.

Autocorrelations are similar to receiver functions from teleseisms in that they can be generated from a single station. The $Ps$ phase and reverberated phases such as $PpPs$ and $PpSs$ commonly appear in the receiver function. These phases are sensitive to velocity jumps and can be used to study subsurface discontinuities. The RF and autocorrelation methods complement each other in determining primary interfaces \citep{delph2019constraining,kim2019groundwater,mroczek2021joint}. ACFs provide independent information on P and S velocities, which are inseparable in RFs. ACFs have difficulties imaging deeper targets such as the Moho, whereas crustal-scale RF studies typically see distinct $Ps$ arrivals from the Moho. This is also true for the BASIN project. While the Moho beneath the studied basins is well resolved by receiver functions \citep{liu2018structure,wang2021urban,ghose2023basin}, we do not see it with autocorrelations. It appears that the Moho is obscured by the strong multiples from the basin bottom.

\section{Conclusions}
Autocorrelating the seismic record at a single station provides an estimate for the zero-offset reflection response. To get rid of the source term, we remove the common mode from the ACF and obtain an enhanced zero-offset section. We derive consistent images of the San Gabriel, Chino, and San Bernardino basins using autocorrelations created from ambient noise and teleseismic waves. The earthquake ACF from a single teleseismic event can also provide good illumination of subsurface. Our images clearly reveal the basin shape that compares well with multiple previous studies covering receiver functions, gravity measurements, and surface wave tomography.  This study provides further constraints on sedimentary basin structures north of Los Angeles for accurately analyzing the resulting seismic amplification.  

\section*{Data and Resources}
The primary data (node data) used in this study are available from the IRIS Data Management Center. The autocorrelations and code underlying this article will be shared on reasonable request to the corresponding author.

\section*{Acknowledgements}
The authors thank all the volunteers who helped with deploying the dense nodal arrays. This research was supported by the National Science Foundation awards 2105358 and 2105320. The BASIN project was partly supported by U.S. Geological Survey awards GS17AP00002 and G19AP00015, and Southern California Earthquake Center awards 18029 and 19033. 

\bibliography{main.bib}
\bibliographystyle{myst}

\clearpage
\begin{table}
\centering
\caption{Teleseismic events selected for computing autocorrelations. \label{events}}
\begin{adjustbox}{max width=\textwidth}
\begin{tabular}{llllllllll}
\toprule
\textbf{Line} & \textbf{\shortstack[l]{Number\\of events}} & \textbf{\shortstack[l]{Selected\\event}} & \textbf{Magnitude} & \textbf{\shortstack[l]{Time\\(yyyy-mm-dd hh:mm:ss)}} & \textbf{\shortstack[l]{Latitude\\(\degree)}} & \textbf{\shortstack[l]{Longitude\\(\degree)}} & \textbf{\shortstack[l]{Depth\\(km)}} & \textbf{\shortstack[l]{Epicentral\\distance (\degree)}} & \textbf{\shortstack[l]{Back\\azimuth (\degree)}} \\
\midrule
SB1 & 15 & Chile & 6.0 & 2019-12-03 08:46:35 (UTC) & -18.504 & -70.576 & 38.0 & 69 & 131 \\
SB2 & 22 & Oregon & 6.2 & 2018-08-22 09:31:45 (UTC) & 43.564 & -127.717 & 10.0 & 12 & 322 \\
SB3 & 22 & Mid-Atlantic Ridge & 6.0 & 2018-07-23 10:35:59 (UTC) & -0.299 & -19.252 & 10.0 & 97 & 85 \\
SB4 & 5 & Argentina & 6.4 & 2017-02-18 12:10:17 (UTC) & -23.861 & -66.659 & 222.0 & 76 & 133 \\
SB5 & 13 & Papua New Guinea & 7.6 & 2019-05-14 12:58:25 (UTC) & -4.051 & 152.597 & 10.0 & 92 & 267 \\
SB6 & 22 & Russia & 6.0 & 2018-08-10 18:12:07 (UTC) & 48.459 & 154.939 & 27.0 & 64 & 312 \\
SG1 & 6 & Solomon Islands & 6.0 & 2017-03-19 15:43:25 (UTC) & -8.136 & 160.754 & 8.4 & 87 & 259 \\
SG2 & 6 & Argentina & 6.4 & 2017-02-18 12:10:17 (UTC) & -23.861 & -66.659 & 222.0 & 76 & 132 \\
SG3 & 13 & Papua New Guinea & 7.6 & 2019-05-14 12:58:25 (UTC) & -4.051 & 152.597 & 10.0 & 92 & 267 \\
SG4 & 13 & Papua New Guinea & 7.6 & 2019-05-14 12:58:25 (UTC) & -4.051 & 152.597 & 10.0 & 92 & 267 \\
\bottomrule
\end{tabular}
\end{adjustbox}
\end{table}

\begin{figure}
    \subfloat[]{\includegraphics[width=1\textwidth]{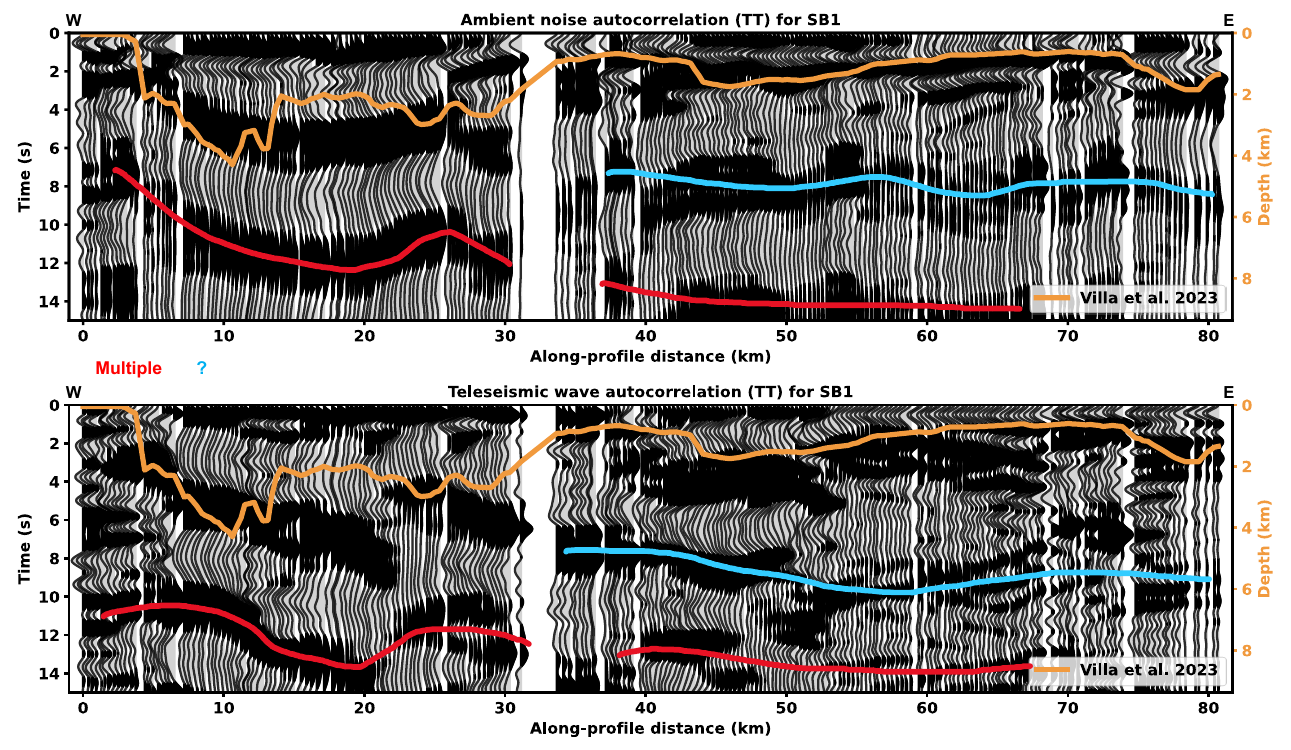}}
    \phantomcaption
\end{figure}
\begin{figure}
    \ContinuedFloat
    \subfloat[]{\includegraphics[width=1\textwidth]{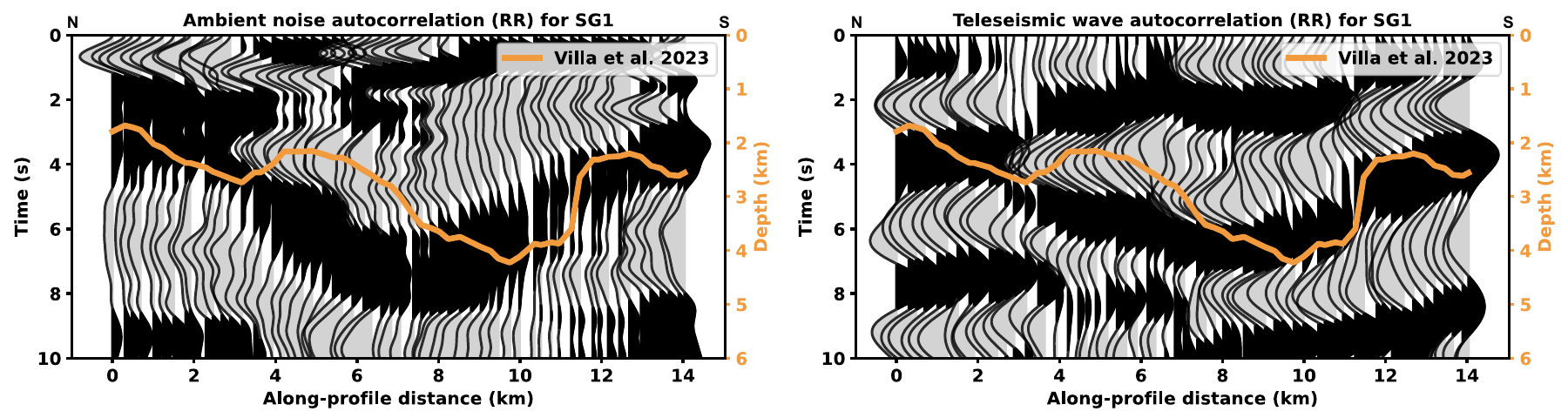}}
    \phantomcaption
\end{figure}
\begin{figure}
    \ContinuedFloat
    \subfloat[]{\includegraphics[width=1\textwidth]{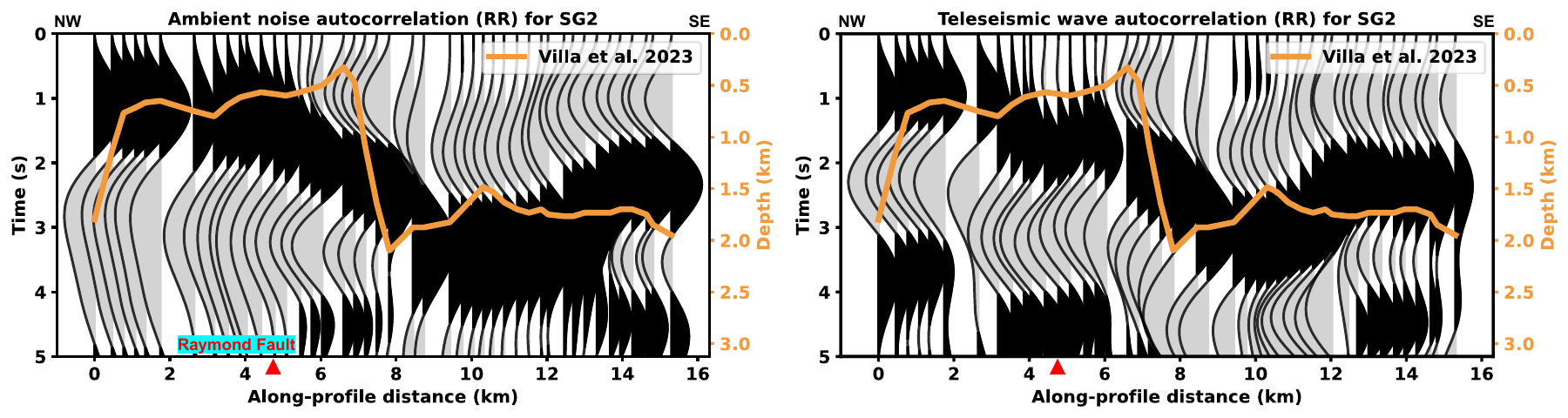}}
    \phantomcaption
\end{figure}
\begin{figure}
    \ContinuedFloat
    \subfloat[]{\includegraphics[width=1\textwidth]{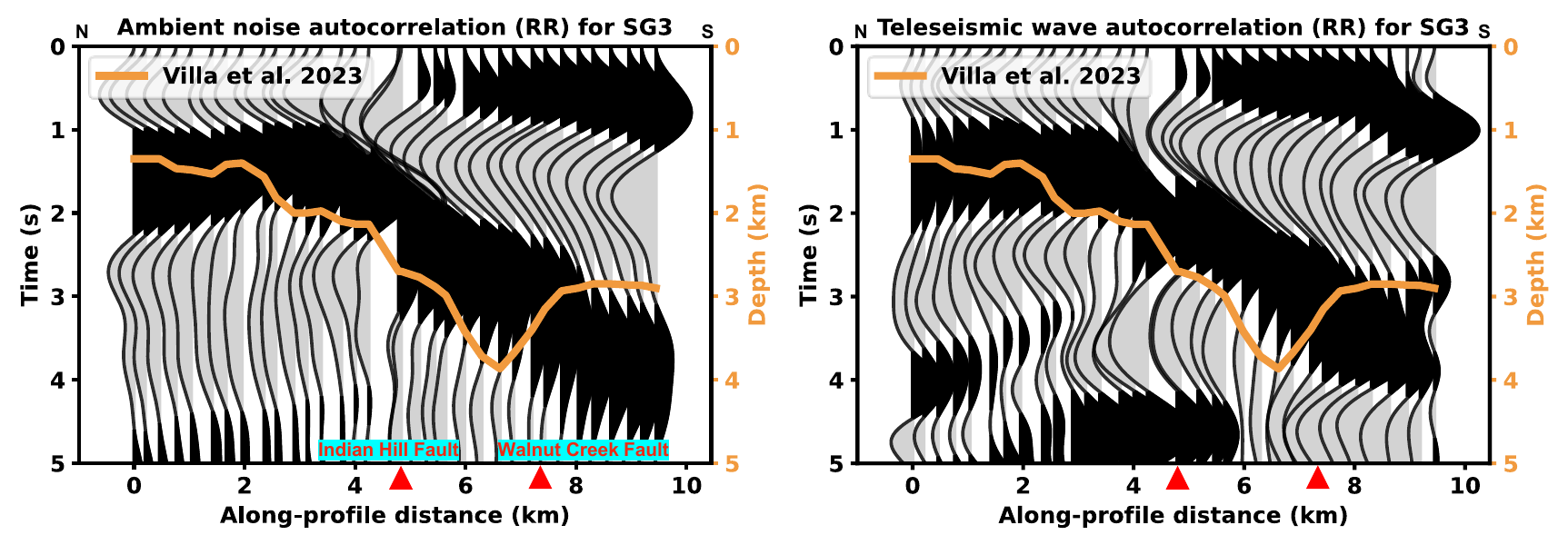}}
    \phantomcaption
\end{figure}
\begin{figure}
    \ContinuedFloat
    \subfloat[]{\includegraphics[width=1\textwidth]{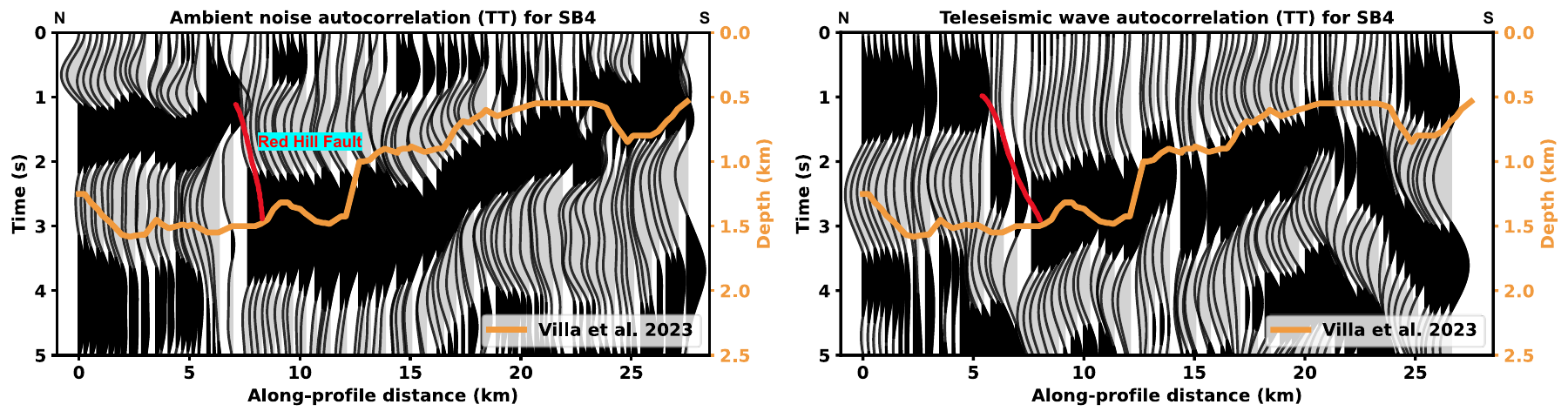}}
    \phantomcaption
\end{figure}
\begin{figure}  
    \ContinuedFloat
    \subfloat[]{\includegraphics[width=1\textwidth]{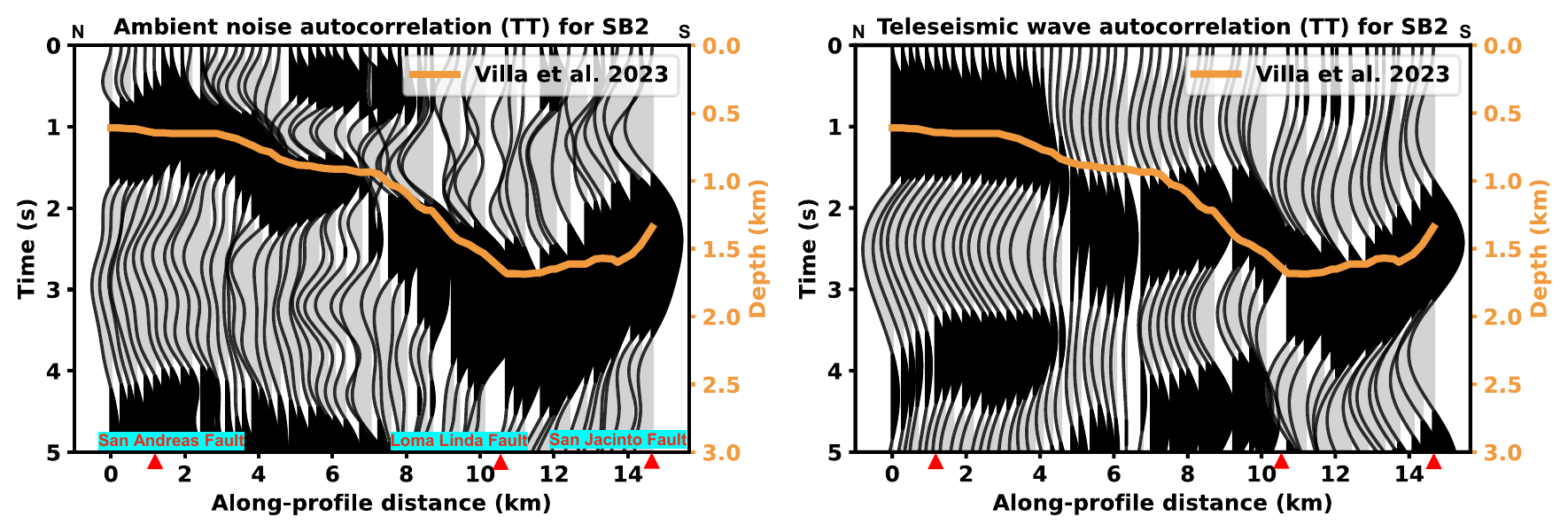}}
    \caption{Imaging results of ambient noise and teleseismic wave autocorrelations for (a) SB1, (b) SG1, (c) SG2, (d) SG3, (e) SB4, and (f) SB2. The yellow line delineates the basin depth derived from \cite{villa2023three}, plotted with the depth axis on the right. Significant faults are annotated and marked with red triangles. In (a), the red line delineates the multiple, and the blue line delineates a geologic feature yet to be investigated. In (e), the red line marks a fault-like feature that is interpreted as the Red Hill Fault.}
    \label{detailedline}
\end{figure}

\label{lastpage}

\end{document}